# Leveraging the Potential of Control-Flow Error Resilient Techniques in Multithreaded Programs

Navid Khoshavi, Mohammad Maghsoudloo, Hamid R. Zarandi

*Abstract*—This paper presents a software-based technique to recover control-flow errors in multithreaded programs. Control-flow error recovery is achieved through inserting additional instructions into multithreaded program at compile time regarding to two dependency graphs. These graphs are extracted to model control-flow and data dependencies among basic blocks and thread interactions between different threads of a program. In order to evaluate the proposed technique, three multithreaded benchmarks quick sort, matrix multiplication and linked list utilized to run on a multi-core processor, and a total of 5000 transient faults has been injected into several executable points of each program. The results show that this technique detects and corrects between 91.9% and 93.8% of the injected faults with acceptable performance and memory overheads.

## 1. Introduction

Improvement in CMOS technology has brought the exponential growth in the number of transistor per chip and decreasing feature size. Reduction in transistor size and voltage levels coupled with increased sensitivity of microprocessors to transient faults [2]. Transient and permanent faults can cause tremendous damage in computer systems, especially in safety-critical systems. Two main groups for designing reliable systems have been proposed in past [16]: 1) Hardware-based approaches, relying on adding custom hardware, and 2) Software-based approaches, relying on exploiting devised software to achieve fault tolerance. One of typical hardware-based solutions is to use of an external hardware like *watchdog (checker) processor* to monitor activities carried out by microprocessors. As soon as any misbehavior is observed, suitable fault containment procedures are activated [1], [4], [12], [13], [15], [16]. Adding redundant hardware for fault tolerance would undermine benefits of modern processors due to area and power impositions. Software-based approaches improve the dependability of processors by acting only on the software, while the underlying hardware remains unchanged. Numerous software-based error detection and error recovery techniques have been devised to assess processor errors and restore the correct system operation in modern processors like multi-core processors [2], [8], [9], [19]. These approaches typically contain error detection and error correction phase. In detection phase, two copies of a same program are executed on the multiprocessor system as separate threads and then their outputs are compared to detect any mismatch. Error correction phase ensures proper supervision of a fault-tolerant algorithm, restoration circuitry and inter-module interconnect.

Some software-based error detection techniques can be applied automatically to the source code of a program, automatically. These techniques that check program execution flow are known as control-flow checking (CFC) techniques. Lots of CFC techniques have been proposed in the past which are mainly based on signature monitoring principle [3], [5], [6], [10], [14], [23]. Some of them are processor-independent and can be applied to any kind of processors and microcontrollers. In these approaches firstly, program code is partitioned into basic blocks and secondly, extra instructions are added to each basic block. Basic block includes a maximal set of ordered non-branching instructions (except in the last instruction) or branch destinations (except in the first instruction) in which the execution begins from the first instruction and terminates at the last instruction [4]. Each basic block is assigned a unique signature. Signatures are calculated at run-time and next compared with the original ones which were calculated at compile time. If any mismatch has observed, an error is detected and reported [3]. It has been shown that between 33% and 77% of transient faults result in control-flow errors, such as possible errors in program counter (PC), address circuits, steering and control logic, or any kind of control memories [21]. A Control-flow Error (CFE) is said to have occurred if the processor executes an incorrect sequence of instructions [3], [27], [28], [29], [30], [31], [32], [33]. Besides these techniques, the error estimation and security techniques have been widely used to handle fault in the current computer architectures [34-38].

Several CFC methods have been designed for detecting each type of CFEs in a program code. However, a few published works have concentrated on CFEs correction [7], [17]. After the CFE is detected, control should be transferred back to the block, where illegal branch has occurred in it. However, correcting the CFE is not sufficient and the program may fail since there may be some data errors generated by the CFEs [7]. Therefore, any data errors caused by CFE should be corrected before correcting the CFE, as well.

In multi-core systems, since all processors share a single view of data and the communication between processors, the method which corrects CFEs and data errors should take into account synchronization and communication dependencies between threads of multithreaded program. Disregarding to thread interactions between different threads by previous CFEs correction techniques caused these methods not be applicable in multithreaded architectures. Therefore, regarding to the importance of handling the CFEs and also disability of the conventional related techniques to be utilized in the modern processors for correcting CFEs, a CFE recovery technique called CRMP (*CFE Recovery for Multithreaded Programs*) is proposed in this paper. Data errors recovery in this technique is based on partial check pointing. Checkpoint-based methods are frequently used to correct data errors [20]. Checkpoints are often saved based on program execution behavior. If any error detected, program is re-executed from the last consistent checkpoint location in the program. However, this may not be possible in safety critical applications, because getting checkpoint, restoring program states and re-executing impose unacceptable area cost and latency [17]. Moreover, consistency of checkpoint view should be guaranteed.

In the proposed technique, only some of variables whose value is changed are checkpointed at the end of each basic block. This approach can considerably reduce the overheads of checkpoint-based methods.

In order to detect CFEs, a signature is used for each basic block, and then some instructions are specified and added at the end of the basic block for calculating and checking signatures at run-time. An error-handler function is prepared at compile time for automatic CFE and data errors correction. This function is implemented regarding to data and control-flow dependencies among basic blocks of a thread and considering synchronization and communication dependencies between threads of multithreaded program. Fault injection is used to evaluate correctability and efficiency of the proposed technique [17]. To evaluate the techniques, three modified multithreaded benchmarks are used and evaluated by GCC, a GNU compiler. A total of 5000 transient faults were injected into the program codes. The experimental results show that between 91.9% and 93.8% of CFEs are detected and corrected by CRMP method without any data errors. Also, the results show that performance and memory overheads of CRMP is considerably less than previous related works like ACCED, CDCC [7], [17].

The structure of this paper is as follows: Section 2 gives related work. Section 3 introduces dependency graph in multithreaded program. Detection and correction phases which are used in proposed technique are described in section 4. Simulation environment and experimental results are presented by section 5. Finally Section 6 concludes the paper.

## 2. Related Work

The Control-Flow Checking by Software Signatures (CFCSS) technique [3] assigns a unique signature to each basic block. A global variable called *G* contains the run-time signature. In absence of error, *G* contains the signature associated to the current basic block. *G* is initialized with the signature of the first block of the program. At the beginning of the basic blocks an additional instruction computes the signature of the destination blocks from the signature of the source block by implementing the XOR function between the signature of the current node and the destination node. If the control can enter from multiple blocks, then an adjusting signature is assigned in each source block and used in the destination block to compute the signature. The CFCSS cannot detect CFEs if multiple nodes share multiple nodes as their destination nodes.

The Yet Another Control-Flow Checking using Assertions (YACCA) technique [21] checks the control flow of programs by using a dedicated global integer variable, which contains the run-time signature associated with the current node in the CFG. A unique signature is assigned to each basic block at the compile time. Two sets of instructions (test and set) are defined and inserted into each basic block. A test instruction controls the signature of the previous basic block and checks if it is permissible, and a set instruction updates the signature, setting it to the correspondent value. Moreover, differently from other approaches, in each basic block the added statements are introduced at the beginning and at the end of the blocks. This approach allows YACCA to detect most of the single inter node CFEs, including the CFEs which are not covered by previous techniques. However, having high performance overhead and high memory overhead are considered two weaknesses of this method. Moreover, CFCSS and YACCA are confined to CFEs detection and do not correct CFEs.

The Automatic Correction of Control-flow Errors (ACCE) technique [17] partitions a program code into functions which include one or more basic blocks. ACCE uses CEDA technique [11] for CFEs detection. After the detection phase, a predefined function called error-handler is automatically executed, and the program control is transferred to the function and then to the basic block in which the illegal jump has been occurred. An extension of ACCE, called Automatic Correction of Control-flow Errors with Duplication (ACCED) [17], is used for data error detection and correction through duplicating instructions. However, the area overhead of ACCED is more than 100 percent which is considerable.

The Control-flow and Data Errors Correction using Data-flow Graph Consideration (CDCC) technique [7] partitions a program code into basic blocks and then insert extra instructions into the program regarding to data-flow graph of program at compile time. Redundant instructions are used to detect and correct CFEs and data errors. In this technique, signature of source basic block (from which the control transferred incorrectly) and the signature of destination one (to which the control transferred illegally) are given to CFE-handler function as inputs and control transferred to the nearest basic block wherein the modified variables between source and destination are initialized. However, this technique cannot apply on systems which need low error correction latency such as hard real-time systems.

## 3. Dependency Graph in Multithreaded Program

A multithreaded program running on multi-core systems has a number of threads that each one has its own control-flow and data-flow. These flows are not independent since inter-thread synchronizations and communications may exist in the program. In order to represent multithreaded program, we present a dependency graph. This graph is composed of connecting graphs of all single threads in the program, using dependency arcs between different threads.

### 3.1 Dependency Graph of a Single Thread (DGST)

A thread is an entity within a process that can be scheduled for execution. All threads of a process share its virtual address space and system resources. In addition, each thread maintains exception handlers, a scheduling priority, thread local storage, a unique thread identifier, and a set of structures the system will use to save the thread context until it is scheduled [2]. However, a thread itself is not a program since it cannot run independently, but it can only run within a program [18]. The dependency graph of a single thread (DGST) is used to represent a single thread in a multithreaded program. DGST consist of a number of Control-flow Graphs (CFGs) and Data-flow Graphs (DFGs). CFG is a graph composed of a set of nodes $V$ and a set of edge $E$, $CFG=\{V,E\}$, where $V=\{N_1, N_2, ..., N_i, ..., N_n\}$ and $E=\{e_1, e_2, ..., e_i, ...,e_n\}$. Each node $N_i$ represents a basic block and each edge $e_i$ represents the branch $br_{i,j}$ from $N_i$ to $N_j$. CFGs are depicted at compile time and represented control conditions and right transmission between basic blocks. DFGs represent data dependencies between basic blocks. The DFG model implicitly assumes the existence of variables, whose values store the information required and generated by the operations. Each variable has a lifetime that is the interval from its birth to its death, where the former is the time at which the value is generated as an output of an operation and the latter is the latest time at which the variable is referenced as an input to operation. Figure 1 shows DFG and CFG generated from control dependency between basic blocks and data dependency among variables in these basic blocks. In this figure, solid and dashed arcs are control and data dependencies, respectively.

### 3.2 Dependency Graph of Multithreaded Program (DGMP)

The Dependency Graph of Multithreaded Program (DGMP) consist of a collection of DGSTs that each represent a single thread, and some special kinds of dependency arcs to model thread interactions between different threads of a program. These dependency arcs are based on: 1) synchronization between thread synchronization statements and 2) communication between shared variables of the program threads.

#### 3.2.1 Synchronization Dependencies

Running each process on its own address space had the advantage of reliability since no process can modify another process's memory [9]. However, all of a process's threads run in the same address space and have unrestricted access to all of the same resources, including memory [8]. While this makes it easy to share data among threads, it also makes it easy for threads to step on each other. Multithreaded programs must be specially programmed to ensure that threads do not step on each other. A section of a code that modifies data structures shared by multiple threads is called a *critical section*. It is important that a critical section should be accessed exclusively by each thread. Synchronize access ensure that only one thread can execute in a critical section at a time. Moreover, synchronization is necessary to avoid race conditions and deadlocks when multiple threads want to access shared resources. Figure 2 shows some additional synchronization arc to model synchronization between threads. As shown in Figure 2, a critical section containing shared data in both basic block $N_3$ of thread $t_1$ and $N_1$ of thread $t_2$ should be access exclusively through the use of synchronization methods.

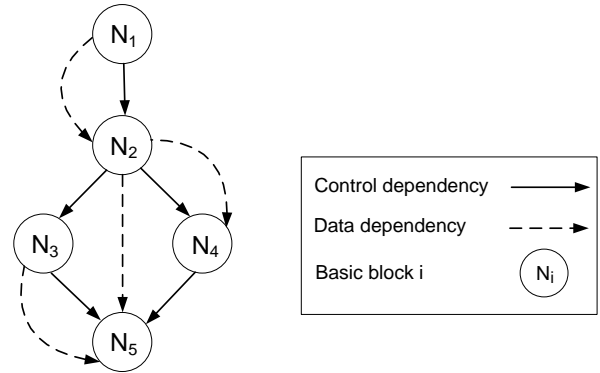

**Figure 1. Dependency graph of a single thread (DGST)**

#### 3.2.2 Communication Dependencies

Communication dependency is used to capture dependency relations between different threads because of inter-thread communication. As shown in Figure 2, node $N_i$ of a thread is directly communication-dependent on node $N_j$ in another thread where $i, j$ are basic block number if the value of a variable computed at $N_i$ has direct influence on the value of a variable computed at $N_j$ through an inter-thread communication. Shared memory is often used to support communication among threads. Communications may occur when two parallel executed threads exchange their data via shared variables.

#### 3.2.3 Constructing the DGMP

To construct the DGMP of a multithreaded program, DGSTs of all program threads using synchronization and communication dependency arcs should be combined. For this purpose, firstly, DGST is constructed with considering control dependency between basic blocks and data dependency among variables of each basic block and secondly DGMP is created based on synchronization and communication dependencies between different threads of multithreaded program as shown in Figure 2. In this figure, bolded dotted and dashed arcs are synchronization and communication dependencies, respectively.

### 4. The Proposed CRMP Technique

CRMP recovers CFEs which are occurred in multithreaded programs in multi-core architectures with using DGMP consideration and partial check pointing. Most of CFEs can be recovered from multithread program by CRMP since DGMP reveals all types of dependencies in the program. So, inter-thread CFEs (CFEs which are occurred from basic block of a thread to basic block of another thread in the same processor) will be regained as well as intra-thread CFEs (CFEs which are occurred between basic blocks of a thread).

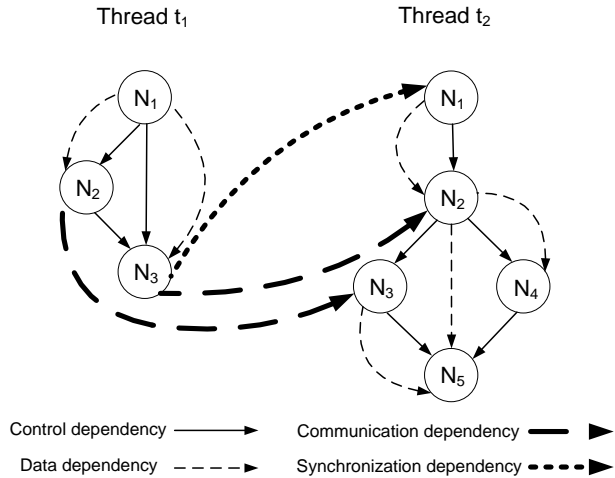

**Figure 2. Dependency graph of multithreaded program (DGMP)**

Two phases are employed to recover CFEs from multithreaded programs.

### 4.1 Control-flow Error Detection in CRMP

As shown in Figure 3, source signature of thread $j$ ($SST_j$) is a shared variable in a shared memory which contains the run-time signature of thread $j$ and continuously updated in executed nodes, where $j$ shows thread number of multithreaded program and finally stored the signature of the basic block in which a CFE has occurred. Each shared variable $SST_j$ which keeps signature of thread $j$ is allowed to be updated only in thread $j$. Another variable $T$ keeps number of thread which is currently executed and it determines source thread in the case of an inter-thread CFE.

$SST_j$ is updated with unique number in each basic block $N_i$ where $i$ shows basic block number of thread $j$, for checking the control-flow of the program. For updating $SST_j$, XOR operation is used. This operation is performed between last $SST_j$ and a number which is calculated at compile time. Under normal execution of the program, $SST_j$ should equal calculated signature in node $N_i$ at compile time. If $SST_j$ contains a number different from the calculated signature in node $N_i$, an error has occurred in the program.

As shown in Figure 4, *CFE1* is an intra-thread CFE because source thread of CFE is itself ($T=1$). If an illegal branch jumped to instructions before checking instructions at the end of basic block (*CFE1* in Figure 4) and control transferred to it illegally, then *CFE1* can be detected by comparing the stored value in the $SST_1$ (as the signature of the node) with another one calculated at compile time.

*CFE2* is an inter-thread CFE (an illegal branch from one basic block of thread $t_2$ to a basic block of thread $t_1$ in the same processor ($T=2$)). These types of CFE can be detected by comparing last updated signature of executed thread with expected value. The signature of last executed basic block of thread $t_1$ is stored in $SST_1$ before switching the CPU from thread $t_1$ to thread $t_2$ and it is equal 1001. The inter-thread *CFE2* can be detected by comparing last updated $SST_1$ (1001) with expected value at the end of basic block.

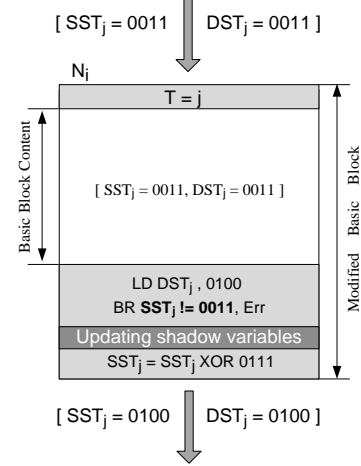

**Figure 3. Basic block scheme in CRMP technique**

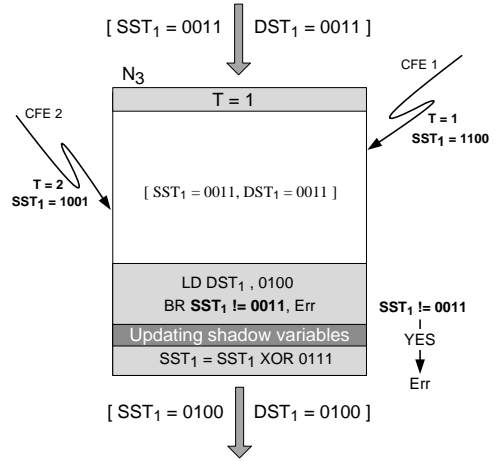

**Figure 4. Intra and inter thread detection**

### 4.2 Control-flow Error Recovery in CRMP

As mentioned in CFE detection section, in the case of a CFE event, data errors may not repair with transferring back to the source block, where illegal branch has occurred in it. So, data errors should be corrected before correcting the CFE. There are two methods for error recovery [8], [9]:
1) Forward Error Recovery: In forward error recovery techniques, the nature of errors and damage caused by faults must be completely and accurately assessed that may not possible.
2) Backward Error Recovery: In backward error recovery techniques, the nature of faults need not be predicted and in the case of error, the process state is restored to previous error-free state.

There are three steps involved in backward error recovery [20], [26]:
1) Periodical check pointing the error-free state,
2) Restoration when a fault occurred,
3) Restart from the restored state.

Depending on the programmer's intervention in check pointing process, backward error recovery can be classified as follows:
1) User-triggered check pointing
2) Transparent check pointing

User triggered check pointing employed where the user has the knowledge of the computation being performed and can decide the location of the checkpoints while transparent check pointing techniques do not require user interaction [26].

User triggered check pointing and backward error recovery are used in CRMP technique and only some of the checkpoint variables are updated at the end of each basic block in which the corresponding original variables has been modified as shown in Figure 3. Through the CRMP, the shadow (checkpoint) variables always contain the true values of the original ones. These values are trustable for correcting the generated data errors in correction phase. Also, it can considerably reduce the overheads of checkpoint-based methods. Information about data dependency among basic blocks variables of each thread is already available at compile time and updating shadows can be specified with respect to it. CRMP recovers CFEs based on type of CFEs.

### 4.2.1 Intra-thread CFE Correction

If an illegal branch jump from one block (*source block*) to another block (*destination block*) in the same thread, original variables which are used in both source and destination block may corrupt [7]. So, source signature of thread *j* ($SST_j$) and destination signature of thread *j* ($DST_j$) are used to keep signature of source and destination block, respectively for correcting corrupted variables in both source and destination block when a CFE occurred in thread *j*.

Shadow (checkpoint) variables always contain the true values of the original ones and they are updated at the end of basic block to store values of variables which are modified in basic block, as shown in Figure 3.

When a CFE is detected through added instructions, control is transferred to CFE-handler function. This function can transfer back the control to the source basic block and restore the original variables with considering information of shadow variables, shared variable *T* and both $SST_j$ and $DST_j$. Figure 5 shows three basic blocks from the set of basic blocks in a program code as well as the DGST generated from data dependencies among variables and control dependency between basic blocks. Suppose that variables *Y* and *Z* are initialized in basic block $N_1$, and variable *X* is initialized in basic block $N_2$. With regards to Figure 5, the time of updating the shadow variables with the original ones can be found. For example, the value of variable *Y* is only changed in basic block $N_1$. Therefore, the shadow of *Y* can be updated only at the end of this basic block instead of updating at the end of the all basic blocks. After a CFE has occurred, it can be detected by checking instructions. The CFE can change the stored values in variables used in computations in the source and the destination. Therefore, it can generate data errors in the results. After detection, control is transferred to CFE-handler function (step 2 in Figure 5). CFE-handler function determines type of CFE by comparing *T* and expected thread number. If an intra-thread CFE had occurred, original variables in destination and source basic block are updated with their shadow ones. Finally, control is transferred to the address of first instruction in the source basic block (step 3 in Figure 5) and the code is re-executed from this point.

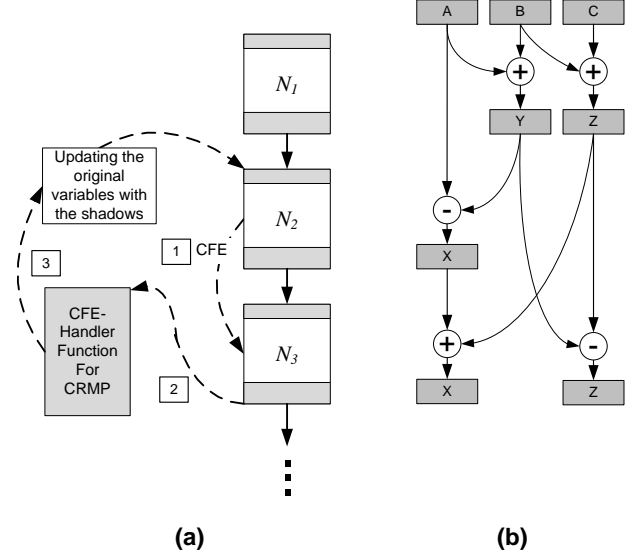

**Figure 5. (a) Intra-thread CFE correction, (b) DFG generated from program code**

Consequently, both of the CFE and the generated data errors can be corrected. For improving the CRMP, the temporal and local variables can be ignored; the act of making shadows for them can be omitted. By this improvement, shadows are only made for global variables which are alternatively used in the program code.

Synchronization dependency among different threads may be caused in two ways [18]: create/join relations, lock/unlock relations.

*Create/join relations.* A main thread can create some more slave threads for performing some subtasks. A function called *pthread_create* has been predefined in most of the multithreaded programming's standards in order to facilitate the explained operation. The main thread should wait for the slave ones until the execution of them run out, if the main thread needs the results which will be provided by the slave ones. This operation has been also simplified through a predefined function called the *pthread_join.* If an illegal branch jumped to the sequential section after basic block containing "*join*" instruction before the synchronization execution, main thread continue with incorrect results which have provided by the slave ones.

*Lock/unlock relations.* As mentioned in dependency graph section, a critical section is a section of code that must be allowed to complete atomically with no interruption that affects its completion. Critical sections are created by locking a lock, manipulating the data, and then releasing the lock. All shared data, data structures which are passed to other threads or can be accessed by multiple threads and global variables must be protected by locks. The mutual exclusion lock is the simplest synchronization variable that provides a single, absolute owner for a section of code. The first thread that locks the mutex gets ownership, and any subsequent attempts to lock it will fail, causing the calling thread to go to sleep. When the owner unlocks it, one of the sleepers will be awakened and it has chance to obtain ownership. *pthread_mutex_lock and pthread_mutex_unlock* functions have been predefined to lock and unlock mutex in most of the multithreaded programming's standards. If a CFE occurred in create/join relations, it will be detected through proposed technique and create/join mechanism corrected by transferring back control to source basic block. However, CFE and data errors will not correct through transferring control to source basic block when CFE occurred in lock/unlock relations as explained below. As shown in Figure 6, both basic block $N_7$ in thread $t_1$ and basic block $N_1$ in thread $t_2$ contain critical section. Basic block $N_7$ in thread $t_1$ is executed and locked the mutex before executing basic block $N_1$ of thread $t_2$. So, thread $t_2$ is executed after executing *unlock()* instruction by thread $t_1$. If a CFE occurred in another critical section (either in thread $t_1$ or another thread at the same processor) and control of program transferred to the basic block $N_7$ of thread $t_1$ (between *lock()* and *unlock()*), the mutex which was locked by previous critical section, is unlocked by executing *unlock()* at the end of critical section of basic block $N_7$. Therefore, thread $t_2$ is allowed to start its code execution before CFE detection in thread $t_1$. This event may cause thread $t_2$ works with dirty shared variables which are corrupted in previous thread.

This type of CFE can be detected when critical section instructions are considered as a basic block and check instructions are putted before *unlock()*, as shown in Figure 7. Communication dependency between basic blocks of different threads means the value of a variable computed at a basic block of a thread has direct influence on the value of a variable computed at basic block of another thread through an inter-thread communication. The CFE time may vary CRMP correction coverage in communication-dependent threads as explained below. Assume an intra-thread CFE occurred in one basic block (assume $N_3$ where can be source or destination basic block) of a thread (assume thread $t_1$) which is communication-dependent to basic block (assume basic block $N_2$) of another thread (assume thread $t_2$) as shown in Figure 2. The terms can be listed based on CFE time as below:
1) $N_2$ of $t_2$ is executed with true shared variables before CFE occurrence in $t_1$.
2) CFE in $t_1$ is detected and corrected before executing $N_2$ of $t_2$.
3) CFE in $t_1$ is detected at the same time $N_2$ is executed in $t_2$.
4) CFE in $t_1$ is detected and corrected after executing $N_2$ of $t_2$ with dirty shared variables.

Outputs of $t_2$ in situation 1, 2 are trustable because shared variables will not be broken when CFE is detected before executing $N_2$ of $t_2$ or when CFE is occurred after executing $N_2$ of $t_2$ with true shared variables. CFE and data errors in situation 3 can be corrected with considering DGMP and signature information of each thread in CFE-handler function. Shared variables in situation 4 are not recoverable, because shadow variables will become dirty if CFE detected after executing $N_2$ of $t_2$ with dirty shared variables. CFE-handler function behavior for correcting control-flow and data errors when communication dependency is exist between basic blocks of different threads can be seen in Figure 8.

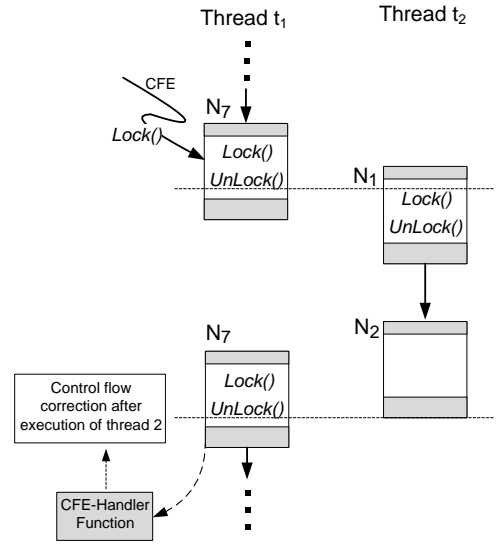

**Figure 6. Wrong intra-thread CFE detection in basic block containing *synchronization* instructions**

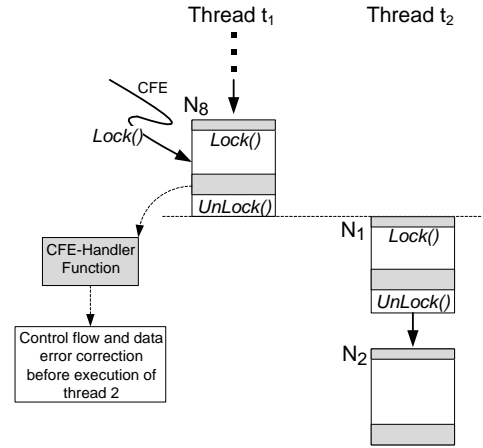

**Figure 7. True intra-thread CFE detection in modified basic block containing *synchronization* instructions**

### 4.2.2 Inter-thread CFE Correction

An inter-thread CFE can be caused by stack pointer faults result in a faulty procedure return. When an illegal branch occurred from $N_2$ of thread $t_1$ to $N_4$ of thread $t_2$, it can be detected by comparing last updated $SST_{destination\ thread}$ with expected value at the end of basic block ($SST_2$ != 0100) as illustrated in Figure 9. After the CFE detection, shared variable $T$ and both signature of thread ($SST_j$ and $DST_j$) are given to CFE-handler function to correct control-flow and data errors. As shown in Figure 8, CFE-handler function determines type of CFE by comparing $T$ and expected thread number. If an inter-thread CFE had occurred, original variables in destination basic block of destination thread are updated with their shadow ones.

Then, CFE-handler found source thread and updated original variables in source basic block of source thread with their shadows and finally control of program returned to source basic block.

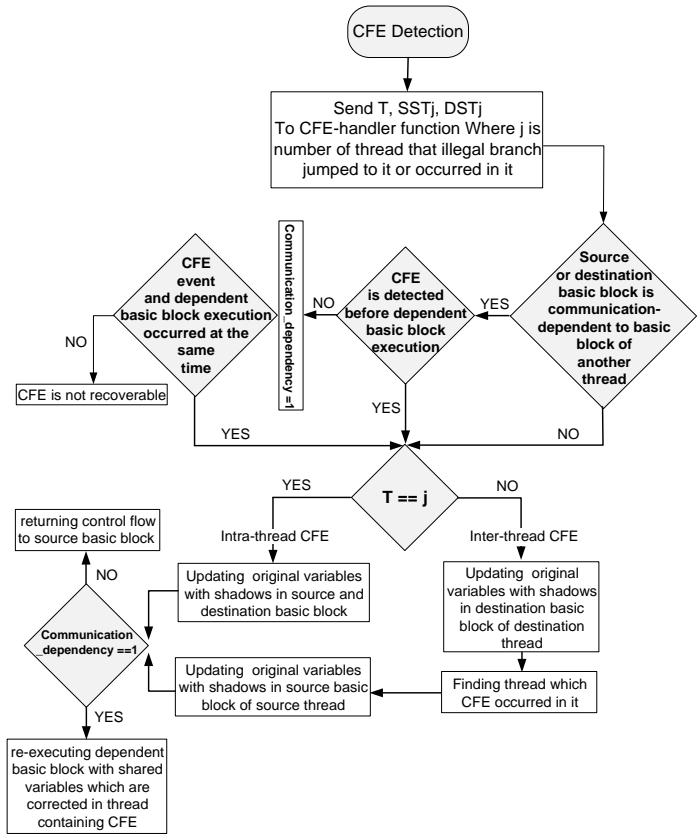

**Figure 8. CFE-handler function pseudo diagram**

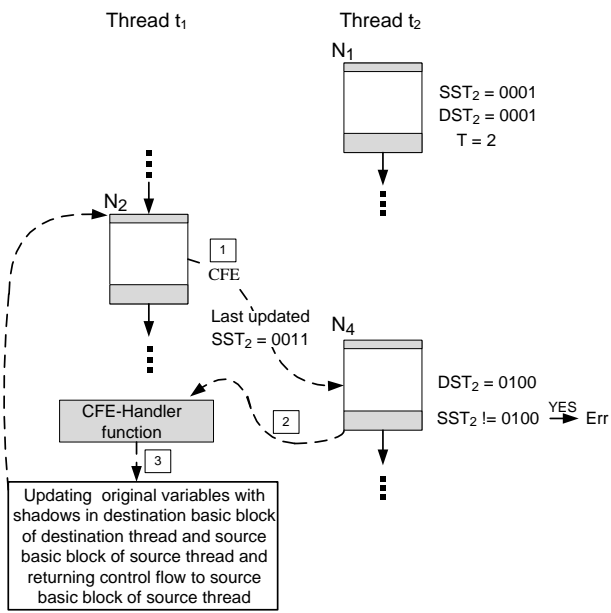

**Figure 9. Inter-thread CFE correction**

## 5. Experimental Results

In order to evaluate the proposed technique, the GCC, a GNU compiler, has been used for running the program on a multi-core system (CPU=i7-740QM, RAM=6GB, OS=Linux Ubuntu10). Three well-known benchmarks have been utilized to execute on multi-core processor system. The main reason of using these benchmarks is to enable us to compare our results with related work. Because these benchmarks are widely used in the related work.

1) Quick Sort (QS): in which the main thread first partitions the 100-elements array of integers into two parts, by performing one round of the quick sort algorithm, then assigns each sub-arrays to a slave thread in order to sort each part separately and simultaneously.
2) Linked List (LL): in which the main thread makes slave threads responsible to build half of the linked list separately and concurrently. After joining the slave threads to the main thread, it connects two splits in order to build a single linked list.
3) Matrix Multiplication (MM): in which the main thread makes slave threads responsible to compute each elements of the product separately and concurrently.

About 5000 transient faults have been injected on several points in the basic blocks of the programs. The considered fault models were [1]:

- Branch insertion: it had occurred when one of the non-control instructions in the program was replaced by a control instruction, and the control instruction always causes a taken branch,
- Branch deletion: it had occurred when one of the control instructions in the program was changed to a NOP instruction.
- Branch target modification: This type of CFE occurred when target address of a control instruction is modified as the result of a fault and the control instruction actually causes a taken branch.

These fault model are used for both intra and inter thread CFEs. On an 80x86 processor, a complete thread switch can be implemented using 12 assembly language instructions. Registers of the processor are pushed onto the stack at the beginning, then stack pointer is changed, and finally they are popped at the end for injecting inter-thread CFEs. Error correction coverage shows the power of the technique for correcting the CFEs. It is said that a CFE has been corrected if the program exits normally with correct output.

Table 1 shows the error correction coverage of the proposed technique in compare to ACCED [17], CDCC [7] and Basic Check-Pointing (BCP) [7] with DGMP consideration. It is found that about 7.6% of the injected faults return correct output without using any technique. As shown in Table 1, about 92.6% of the injected faults return correct output with CRMP. On average, about 7% of the faults cause segmentation faults for the suggested techniques.

**Table 1. Experimental fault injection results for multithreaded benchmarks**

| Benchmarks | Original | | ACCED [17] | | BCP [7] | | CDCC [7] | | CRMP | |
|---|---|---|---|---|---|---|---|---|---|---|
| | *Wrong Results* | *Correct Results* | *Wrong Results* | *Correct Results* | *Wrong Results* | *Correct Results* | *Wrong Results* | *Correct Results* | *Wrong Results* | *Correct Results* |
| | % | % | % | % | % | % | % | % | % | % |
| QS | 92.4 | 7.6 | 5.5 | 94.5 | 8.8 | 91.2 | 7.5 | 92.5 | 7.9 | 92.1 |
| MM | 93.7 | 6.3 | 4.7 | 95.3 | 6.6 | 93.4 | 5.2 | 94.8 | 6.2 | 93.8 |
| LL | 91.3 | 8.7 | 6.1 | 93.9 | 8.4 | 91.6 | 8.5 | 91.5 | 8.1 | 91.9 |
| Average | 92.4 | **7.6** | 5.1 | **94.9** | 7.7 | **92.3** | 7.1 | **92.9** | 7.4 | **92.6** |

Segmentation faults are generated if the illegal branches jump to signature update statements or a *push* or *pop* instruction re-execute due to CFE correction.

CFEs which are not corrected can be classified as follows:
1) Not detected CFEs are included:
   - Inter-thread CFE to basic block of another thread where if CPU switch to that thread, start from that basic block
   - Intra-node CFE (an illegal movement within a basic block)
   - Ilegal branch to CFE-handler function
2) Detected CFEs but not corrected are included:
   - Ilegal jump to *critical instructions*
   - CFEs in communication-dependent threads when executed thread work with shared data which are corrupted in previous thread

Critical instructions of CRMP where if an illegal branch jumped to it, CFE will detect but not correct are more than critical instructions of ACCED and CDCC. These instructions include signature update statement, shadow variables and *T* update statement. So, correction coverage of CRMP is slightly fewer than correction coverage of ACCED and CDCC.

*CFE detection latency* is equal to the time between fault occurrence and the time at which the CFE caused by the injected fault is detected. CFE detection latency of CRMP is slightly more than other techniques because it only uses checking instruction at the end of each basic block. Figure 10 shows the CFE detection latency observed during fault injection in terms of the number of instructions executed before the CFE is detected.

*CFE correction latency* of CRMP consists of:
- CFE detection latency
- Latency of the CFE-handler function execution
- Latency of updating corrupted variables with shadow ones
- Latency of control transferring to source basic block

Figure 11 shows the CFE correction latency observed during fault injection in terms of the number of instructions executed for restoring the CFE and correcting data errors.

Figure 12 illustrates the comparison among *performance overhead* percentages of the programs due to applying the methods. The performance overhead (cost) that reveals the amount of performance degradation due to methods operation, and is estimated as follow:

Performance Cost=Δ Performance / Initial Performance (1)

Figure 13 illustrates the comparison among *memory overhead* percentages of the programs due to applying the methods. The memory overhead (cost) consists of the set of instructions which are added at the beginning and at the end of the basic blocks and the other set added for implementing the CFE-handler function. The memory cost of the methods is calculated as below:

Memory Cost=Δ Memory Usage / Initial Memory Usage (2)

The memory (performance) overhead of the ACCED is comparatively (about 100%) higher than the proposed technique because of adding duplicated instructions and executing the set of instructions used for comparing the results to obtain correct output.

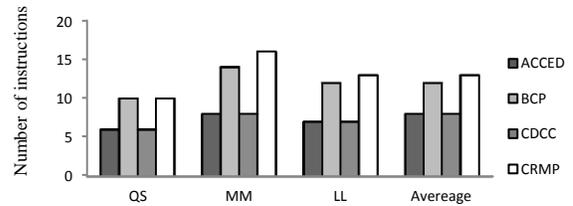

**Figure 10. Comparison of CFE detection latency in different methods**

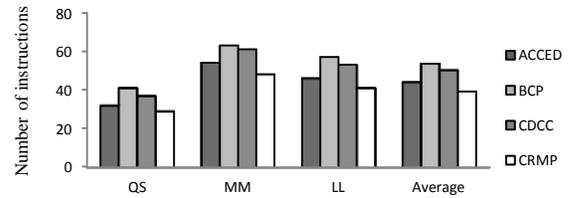

**Figure 11. Comparison of CFE correction latency in different methods**

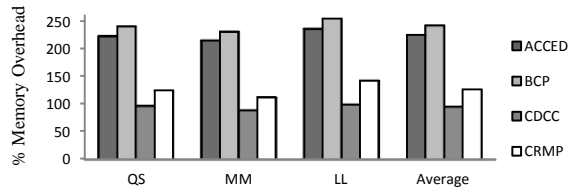

**Figure 13. Comparison of memory overhead in different methods**

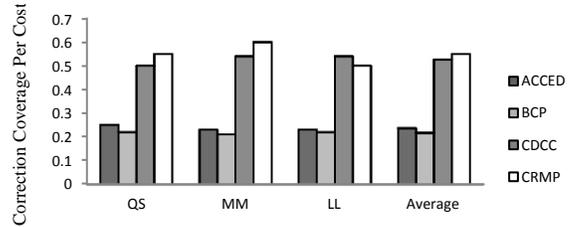

**Figure 14. Comparison of efficiency in different methods**

**Table 2. Comparison of the proposed technique with some of the previous CFC methods**

| CFC method | Error detection coverage (%) | Error correction coverage (%) | Memory overhead (%) | Performance overhead (%) | Method efficiency | Correctability | Applicability in multithreaded architectures |
|---|---|---|---|---|---|---|---|
| YACCA[21] | 21.1-56.0 | 0 | 191.0-496.0 | 110.0-354.0 | 0.12-0.18 | NO | NO |
| CFCSS[3] | 28.8-41.0 | 0 | 26.6-63.6 | 16.2-69.2 | 0.14-0.22 | NO | NO |
| ACCED[17] | 94.0-97.5 | 93.9-95.3 | 215.5-236.0 | 149.5-189.8 | 0.23-0.25 | YES | NO |
| CDCC[7] | 92.3-95.7 | 91.5-94.8 | 89.0-99.4 | 69.9-89.5 | 0.50-0.54 | YES | NO |
| BCP [7] | 91.5-94.8 | 91.2-93.4 | 230.8-254.3 | 153.0-199.0 | 0.21-0.22 | YES | NO |
| CRMP | 92.0-95.0 | 91.9-93.8 | 112.0-142.5 | 38.0-43.5 | **0.50-0.60** | YES | **YES** |

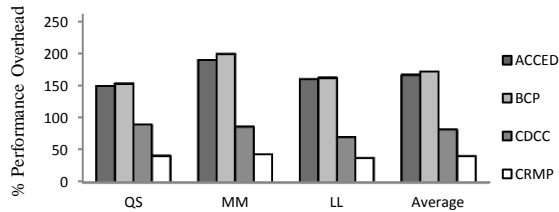

**Figure 12. Comparison of performance overhead in different methods**

The overall cost is differently calculated depended on the type of the applications on which the techniques are applied. Two impact factors have been defined for costs: α and β. The impact factor of the performance cost in the system is estimated by α, and the impact factor of the memory cost is estimated by β. In each system, in order to balance α, β, sum of these impact factors should be equal to one (β=1-α).

In some applications the performance cost is more important than the memory one. In these situations, the value of α should be rather than β. On the contrary, if the importance of the memory cost is higher than the performance one, the value of β should be rather than α. The cost is estimated as follow:

Total Cost = α × Performance Cost + β × Memory Cost (3)

Without loss of generality, α = β = 0.5, this means that the importance of the performance and memory cost are equal in our estimation.

In order to estimate the efficiency of the methods, a metric (which is called *Correction Coverage per Cost or CCC*) has been defined, as below:

CCC = Correction Coverage / Cost (4)

where *Correction coverage* (Error correction coverage of the methods) is the percentage of the injected CFEs which are corrected by the technique.

According to Figure 14, the *CCC* of the proposed technique is more than twice the *CCC* of the ACCED, and it means that the proposed technique is more efficient than the ACCED [17]. The memory overhead of CDCC [7] is lower than CRMP and its performance overhead is fair. So, efficiency of CDCC is near to efficiency of CRMP.

Performance overhead of the *matrix multiplication* and the memory overhead of the *linked list* are totally higher than the others. The matrix multiplication has many computational operations, and the basic blocks of its code are larger than the other ones. Consequently, for correcting a CFE, the instructions which should be re-executed is more than other benchmarks. In contrast to matrix multiplication, linked list program has the fewest computational operations, and the number of the basic blocks separated in the linked list is more than the other benchmarks. Therefore, the total of added instructions at the end of the basic blocks is a large number in compare to the others.

A comprehensive comparison among proposed technique and previous CFC techniques are reported in Table 2. As reported in Table 2, the three advantages of the proposed technique are acceptable overheads, preferable efficiency and correctability in multithreaded programs.

## 6. Conclusions

In this paper, a software technique to detect and correct CFEs in multithreaded programs was proposed. This technique was implemented through considering control and data dependency in DGST beside synchronization and communication dependency in DGMP at compile time. Also, proposed technique corrects data errors generated by CFEs that can cause considerable corruptions in the systems (especially in the safety critical applications). Fault injection experiments showed that the proposed technique, when applied on the programs, produce correct results in over 92.6% of the cases. The latency and the additional memory required for correcting the CFEs and the data errors are considerably less than the duplication based and checkpoint based methods which have been recently published. A metric for estimating efficiency of the techniques was defined, and it was shown that the proposed technique is more efficient than the duplication-based and checkpoint-based methods.